# Exoplanet Exploration Program Analysis Group (ExoPAG) Report to Paul Hertz Regarding Large Mission Concepts to Study for the 2020 Decadal Survey

October 6, 2015

B. Scott Gaudi (OSU), Eric Agol (Washington), Daniel Apai (Arizona) Eduardo Bendek (NASA Ames), Alan Boss (Carnegie DTM), James B. Breckinridge (Caltech), David R. Ciardi (NExScI/IPAC), Nicolas B. Cowan (McGill), William C. Danchi (NASA GSPC), Shawn Domagal-Goldman (NASA GSFC), Jonathan J. Fortney (UCSC), Thomas P. Greene (NASA Ames), Lisa Kaltenegger (Cornell), (James F. Kasting (Penn State), David T. Leisawitz (NASA GSFC), Alain Leger (Paris U.), Charles F. Lille (Lille Consulting), Douglas P. Lisman (JPL), Amy S. Lo (Northrup Grumman), Fabian Malbet (IPAG), Avi M. Mandell (NASA GSFC), Victoria S. Meadows (Washington), Bertrand Mennesson (JPL), Bijan Nemati (JPL), Peter P. Plavchan (Missouri State), Stephen A. Rinehart (NASA GSFC), Aki Roberge (NASA GSFC), Eugene Serabyn (JPL), Stuart B. Shaklan (JPL), Michael Shao (JPL), Karl R. Stapelfeldt (NASA GSFC), Christopher C. Stark (STScI), Mark Swain (JPL), Stuart F. Taylor (Participation Worldscope), Margaret C. Turnbull (SETI), Neal J. Turner (JPL), Slava G. Turyshev (JPL), Stephen C. Unwin (JPL), Lucianne M. Walkowicz (Adler Planetarium), on behalf of the ExoPAG

## Joint PAG Executive Summary:

### The PAGs Concur That All Four Mission Concepts Should Be Studied.

This is a joint summary of the reports from the three Astrophysics Program Analysis Groups (PAGs) in response to the charge given to the PAG Executive Committees by the Astrophysics Division Director, Paul Hertz, in the white paper[1] "Planning for the 2020 Decadal Survey", issued January 4, 2015. This joint executive summary is common to all three PAG reports, and contains points of consensus across all three PAGs, achieved through extensive discussion and vetting within and between our respective communities. Additional information and findings specific to the individual PAG activities related to this charge are reported separately in the remainder of the individual reports. These additional findings are not necessarily in contradiction to material in the other reports, but rather generally focus on findings specific to the individual PAGs.

The PAGs concur that all four large mission concepts identified in the white paper as candidates for mission concept maturation prior to the 2020 Decadal Survey should be studied in detail. These include the Far-IR Surveyor, the Habitable-Exoplanet Imaging Mission, the UV/Optical/IR Surveyor, and the X-ray Surveyor. Other flagship mission concepts were considered, but none achieved sufficiently broad community support to be elevated to the level of these four primary candidate missions.

This finding is predicated upon assumptions outlined in the white paper and subsequent charge, namely that 1) major development of future large flagship missions under consideration are to follow the implementation phases of the James Webb Space Telescope (JWST) and the Wide-Field InfraRed Survey Telescope (WFIRST); 2) NASA will partner with the European Space Agency on its L3 Gravitational Wave Surveyor, participate in preparatory studies leading to this observatory, and conduct the necessary technology development and other activities leading to the L3 mission, including preparations that will be needed for the 2020 decadal review; and 3) that the Inflation Probe be classified as a probe-class mission to be developed according to the technology and mission planning recommendations in the 2010 Decadal Survey report[2]. The Physics of the Cosmos PAG (PhysPAG) sought input on the mission size category for this mission and finds that it is appropriately classified as a Probe-class mission. If these key assumptions were to change, this PAG finding would need to be re-evaluated in light of the changes.

The PAGs find that there is strong community support for the second phase of this activity - maturation of the four proposed mission concept studies. The PAGs believe that these concept studies should be conducted by scientists and technical experts assigned to the respective Science and Technology Definition Teams (STDTs). The PAGs find that the community is concerned about the composition of these STDTs and that there is strong consensus that all of the STDTs contain broad and interdisciplinary representation of the science community. The PAGs also find that the community expects cross-STDT cooperation and exchange of information whenever possible to facilitate the sharing of expertise, especially in the case of the UVOIR Surveyor and the Habitable-Exoplanet Imaging Mission, which share some science goals and technological needs. The PAGs concur that a free and open process should be used to competitively select the STDTs.

Finally, the PAGs find that there is community support for a line of probe-class missions within the Astrophysics Division mission portfolio. The PAGs would be willing to collect further input on probe missions from the community as a following strategic planning charge if asked to do so by the Astrophysics Division Director.

| Far-IR Surveyor | Habitable-Exoplanet Imaging Mission | Large UV/Optical/IR Surveyor | X-ray Surveyor |
|---|---|---|---|
| **Primary science goals:**<br><br>• History of energy release in galaxies: formation of stars, and growth of black holes.<br>• Rise of the first heavy elements from primordial gas.<br>• Formation of planetary systems and habitable planets.<br><br>**Measurement Requirements:**<br><br>• Spectral-line sensitivity better than $10^{-20}$ Wm$^{-2}$ in the 25-500 µm band. (5 sigma, 1h)<br>• Imaging spectroscopy at R~500 over tens of square degrees.<br>• R~10,000 imaging spectroscopy of in thousands of z<1 galaxies and protoplanetary disks.<br>• High-spectral-resolution capabilities desired for Galactic star-forming systems and the Galactic Center.<br><br>**Architecture and Orbit:**<br><br>• Complete spectroscopic coverage at R~500 from 25-500 µm.<br>• Monolithic telescope cooled to <4 K, diameter ~5 m.<br>• Field of View = 1 deg at 500 µm<br>• R~10,000 mode via etalon insert.<br>• Background limited detector arrays with few x $10^5$ pixels, likely at T<0.1 K.<br>• Mission: 5 years+ in L2 halo orbit.<br>• High-resolution (heterodyne) spectroscopy under study, possibly for warm phase. | **Primary science goals:**<br><br>• Direct imaging of Earth analogs, search for potential habitability.<br>• Cosmic origins science capabilities considered baseline.<br><br>**Measurement Requirements:**<br><br>ExoEarth detection and characterization requirements:<br><br>• ~$10^{-10}$ contrast<br>• Coronagraph and/or starshade<br>• Optical and near-IR camera for planet detection and characterization<br>• IFU, R>70 spectrum of 30 mag exoplanet<br>• 1" radius FOV<br><br>Cosmic Origins Science requirements:<br><br>• UV-capable telescope/instrument suite: properties and wavelength range to be determined.<br>• Enable constraints on the high-energy radiation environment of planets.<br><br>Possible instrument for spectroscopic characterization of transiting planets<br><br>**Architecture and Orbit:**<br><br>• Aperture: <~8m likely<br>• Monolithic or segmented primary<br>• Optimized for exoplanet direct imaging,<br>• Orbit: L2 or Earth-trailing likely. | **Primary science goals:**<br><br>• Direct imaging of Earth analogs, search for biosignatures.<br>• Broad range of cosmic origins science<br><br>**Measurement Requirements:**<br><br>Cosmic Origins Science requirements:<br><br>• HST-like wavelength sensitivity (FUV to Near-IR)<br>• Suite of imagers/spectrographs, properties to be determined.<br><br>ExoEarth detection and characterization requirements:<br><br>• ~$10^{-10}$ contrast<br>• Coronagraph (likely), perhaps with a starshade<br>• Optical and near-IR camera for planet detection and characterization.<br>• IFU, R>70 spectrum of 30 mag exoplanet<br>• 1" radius FOV<br>• Possible instrument for spectroscopic characterization of transiting planets.<br><br>**Architecture and Orbit:**<br><br>• Aperture: ~8-16m likely<br>• Likely segmented, obscured primary.<br>• Orbit: L2 likely | **Primary science goals:**<br><br>• Origin and growth of the first supermassive black holes.<br>• Co-evolution of black holes, galaxies, and cosmic structure.<br>• Physics of accretion, particle acceleration, and cosmic plasmas.<br><br>**Measurement Requirements:**<br><br>• Chandra-like (0.5") angular resolution<br>• Detection sensitivity ~ 3 x $10^{-19}$ erg cm$^{-2}$ s$^{-1}$<br>• Spectral resolving power: R>3000 @ 1 keV; R~1200 @ 6 keV<br><br>**Architecture and Orbit:**<br><br>• Effective area ~3 m$^2$<br>• Sub-arcsecond angular resolution<br>• High-resolution spectroscopy (R ~ few x $10^3$) over broad band via micro-calorimeter & grating spectrometer instruments<br>• FOV ≳ 5′<br>• Energy range ~0.1-10 keV<br>• Orbit: L2 likely |

Table 1: Notional Mission Parameters. These are the notional parameters of the four missions, developed through coordinated discussions with and between the three PAGs. We emphasize that these parameters are notional: they are not meant to provide definitive or restrictive specifications for range of possible range of architectures to be studied by the STDTs. We encourage the STDTs to consider architectures and parameters outside of those indicated here, in order to explore the full range of science goals, and maximize the science achievable by these missions given constraints on the cost, schedule, and technological readiness. Note by ``Earth analogs'' above we mean, very roughly, terrestrial (i.e., primarily rocky) planets with thin atmospheres in the habitable zones of their parent star.

## 1. Introduction to the Charge

On January 4, 2015, Paul Hertz (director of NASA's Astrophysics Division) released a white paper[1] entitled "Planning for the 2020 Decadal Survey." This white paper issues a charge to the three Program Analysis Groups (PAGs: the Cosmic Origins PAG, Exoplanet Exploration PAG, and Physics of the Cosmos PAG) to identify a small (3-4) set of large mission concepts to be studied by Science and Technology Definition Teams (STDTs) in advance of, and in preparation for, the 2020 Astrophysics decadal survey. For the purpose of this charge, "large" was defined as over ~$1B. The goals for each of these teams will be, at a minimum, to develop the science case for the mission, construct a design reference mission with strawman payload, identify technology development needs for the mission, and provide a cost assessment for the mission.

The white paper identified four potential mission concepts, noting that the PAGs should feel free to add, subtract, or combine these mission concepts, keeping in mind the final set of identified missions should be kept small. These missions were:

- **Far IR Surveyor** – The Astrophysics Visionary Roadmap identifies a Far-IR Surveyor with improvements in sensitivity, spectroscopy, and angular resolution.
- **Habitable-Exoplanet Imaging Mission** – The 2010 Decadal Survey[2] recommends that a habitable-exoplanet imaging mission be studied in time for consideration by the 2020 decadal survey.
- **Large UV/Optical/IR Surveyor**[a] –The Astrophysics Visionary Roadmap identifies a Large UV/Optical/IR Surveyor with improvements in sensitivity, spectroscopy, high contrast imaging, astrometry, angular resolution and/or wavelength coverage. The 2010 Decadal Survey recommends that NASA prepare for a UV mission to be considered by the 2020 decadal survey.
- **X-ray Surveyor** – The Astrophysics Visionary Roadmap identifies an X-ray Surveyor with improvements in sensitivity, spectroscopy, and angular resolution.

As described, three of these mission concepts (the Far-IR Surveyor[b], the Large UV/Optical/IR Surveyor, and the X-Ray Surveyor), were drawn from the NASA Astrophysics Roadmap[3] "Enduring Quests, Daring Visions NASA Astrophysics in the Next Three Decades". The fourth (the Habitable-Exoplanet Imaging Mission) was drawn from the 2010 "New Worlds, New Horizons" Decadal Survey Report[2].

However, we note that none of these four mission concepts, as presented in either NASA Roadmap, or in the decadal survey report, were well defined in terms of their specific science capabilities, architectures, technology requirements, required ground-based supporting observations, or costs. Indeed, it is precisely the goal of the STDTs that will be convened to study these missions to define these properties. Nevertheless, in our initial discussions, we quickly realized that coordination within and between the PAGs would be very difficult without any sort of (even qualitative) specification of the range of probable architectures for each mission. Similarly, it would be difficult to have any discussion of science applications without agreeing upon at least some (qualitative) capabilities, and thus defining approximate ranges of the possible architectures of these missions. Without being able to specify even rough science applications, we found that it was very difficult to identify which set of missions to study, and thus found it nearly impossible to respond to our charge. As a result, during a joint meeting at the Space

---

a The communities most interested in this mission view this as a telescope with a large aperture, so the acronym LUVOIR (Large UV/Optical/IR Surveyor) will be used to refer to this mission in this document, as it was called in the NASA Astrophysics Roadmap[3].
b The NASA Astrophysics roadmap used the title "Surveyor" to distinguish missions envisioned to be developed in the 2020s from those developed in the 2030s and beyond. However, this title was not meant to imply that these missions were primarily focused on surveys.

Telescope Science Institute on March 19, 2015, members of the Executive Committees of the three PAGs wrote down a straw man list of (a range of) potential mission architectures. We have since refined and augmented this list during subsequent discussions. This list is presented in Table 1. We emphasize that the parameters in this table are *strictly notional*: they are not meant to provide definitive or restrictive specifications for the range of possible range of architectures to be studied by the STDTs. We encourage the STDT to consider architectures and parameters outside of those indicated here, in order to explore the full range of science goals, and maximize the science achievable by these missions given constraints on costs, schedule, and technological readiness.

Ultimately, after considerable discussion within and between the PAGs, our final and most important conclusion is that **all three PAGs concur that all four large mission concepts listed in white paper should be studied in detail, and that no missions should be subtracted, added, or merged.** We note that this conclusion is subject to the boundary conditions specified in the white paper and subsequent charge, and discussed in more detail in the PhysPAG report, namely that 1) major development of future large flagship missions under consideration are to follow the implementation phases of JWST[4] and WFIRST[5]; 2) NASA will partner with the European Space Agency on its L3 Gravitational Wave Surveyor, participate in preparatory studies leading to this observatory, and conduct the necessary technology development and other activities leading to the L3 mission, including preparations that will be needed for the 2020 decadal review; and 3) that the Inflation Probe is indeed classified as a probe-class mission to be developed according to the technology and mission planning recommendations in the 2010 Decadal Survey report.

## 2. Potential Exoplanet Science Applications of the Four Proposed Missions

The science goals pursued by the exoplanet community have relevance to all of the top-level questions contained in the 2014 NASA Science plan[6]. The primary goals of the field of exoplanets in the next 10-20 years include: understanding of the diversity of planets and planetary systems in our local part of the universe; understanding of the properties of these worlds and how these properties are affected by interactions with their parent stars; placing the Earth, the Sun, and the rest of our solar system in the context of observations of other planetary systems; and the search for habitable worlds and biosignatures on these worlds. The community feels that addressing all of these questions is worthy of flagship-level investment of resources. The ExoPAG felt that all four proposed mission concepts address, to varying extents, these science goals. Therefore, we recommend STDTs be convened to study each of these flagship missions. The exoplanet science rationale for each mission is outlined below. We defer to the COPAG and PhysPAG reports for the non-exoplanet science applications of these missions, but note that, in particular, there is potentially substantial synergy between the exoplanet and cosmic origins science goals for the nominal architectures of the Habitable-Exoplanet Imaging and LUVOIR missions. In particular, **the COPAG and ExoPAG concur that, in order to ensure broad support for the HabEx and LUVOIR missions within both the exoplanet and cosmic origins communities, significant science capabilities in both topical areas must be baselined for these missions.**

### 2.1 The Far-IR Surveyor

A Far-IR Surveyor would also have significant exoplanet science applications. Possible realizations of such a mission have been previously explored in the NASA Astrophysics Roadmap[3] as well as more specific concept studies. We provide a notional range of possible mission parameters for the Far-IR Surveyor in Table 1; these were adopted from the Far-IR Workshop Final report[7]. The majority opinion from this workshop was that a monolithic, single-dish design (rather than an interferometer) for the Far-IR Surveyor would be the best architecture to address the primary cosmic origins science questions over the next two decades.

Direct observations of planets with a single-dish Far-IR Surveyor will not be possible, due to the limited angular resolution at these long wavelengths. An interferometric Far-IR Surveyor with sufficient spatial resolution could potentially observe some giant exoplanets, but will not have the sensitivity to observe smaller ones due to their extreme faintness at these long wavelengths.

Nevertheless, there is valuable information on planet formation that may be obtained with a Far-IR Surveyor. This wavelength region contains strong transitions of many gases and solid materials critical for studying abundances and chemistry in protoplanetary disks, making it a key wavelength range for learning about planet formation processes. Lines of both water vapor and ice appear at mid- to far-IR wavelengths, permitting a probe of the abundance and distribution of material vital for formation of giant planet cores and the emergence of habitable conditions on rocky worlds. Lines of many other organic molecules lie in the far-IR as well. Furthermore, molecular hydrogen should make up the bulk of the mass in protoplanetary disks, but is notoriously hard to observe. The rotational lines of HD in the far-IR are a tracer of $H_2$, providing valuable information for constraining total disk masses. These planet-forming materials' spatial distributions can be inferred by combining unresolved measurements with modeling, or directly measured at the sub-arcsecond spatial resolutions possible with an interferometric Far-IR Surveyor.

Turning to the later stages of planet formation, when rocky planets complete their formation and the architectures of planetary systems may be rearranged, this mission could take surveys for cold Kuiper Belt dust in debris disks to the next level. There is a strong need to expand our survey volumes and push to lower dust levels. It may also be possible to discover additional targets for follow-up at sub-mm to mm wavelengths with the Atacama Large Millimeter/submillimeter Array (ALMA). Sensitive far-IR spectroscopy capability could grow the nascent field of debris disk gas and dust composition, probing the make-up of young extrasolar comets and asteroids, which are the building blocks of terrestrial exoplanets and their atmospheres.

The NASA Astrophysics Roadmap envisaged a space-based far-IR interferometer as a valuable technological precursor to more challenging future interferometers operating at shorter wavelengths. An example of particular interest to the exoplanets community is a mid-IR interferometer for direct exoplanet characterization at thermal wavelengths (see Section 3 on a terrestrial planet imaging interferometer). A single-aperture Far-IR Surveyor (as currently envisioned) will not address the long-term development of space-based interferometry. Significant investment in space-based interferometry technology in the coming decades will be needed so that a mid-IR mission capable of exoplanet characterization could fly in the foreseeable future.

## 2.2 The Habitable-Exoplanet Imaging Mission (HabEx)

A Habitable-Exoplanet Imaging Mission, or HabEx, was specifically called out as a potential mission for study in the 2010 "New Worlds, New Horizons" Decadal Survey Report. One possible realization of such a mission has been previously explored in the THEIA mission concept report[8] submitted to the 2010 Decadal Survey. Furthermore, a HabEx mission study could leverage much of the work presented in the smaller, probe-scale Exo-C[9] and Exo-S[10] direct imaging mission concept reports. We provide a notational range of possible mission architectures for HabEx in Table 1. HabEx would present different technical challenges than those of the LUVOIR discussed below, and yet could pursue many of the science goals of that mission. Specifically, the ExoPAG community feels such a mission would be worthy of consideration for flagship-class support if it could:

1) Search for and characterize potentially habitable[c] worlds.
   a) Search for – via direct detection of reflected starlight – Earth-sized planets in the habitable zones of other stars; with a searchable sample size sufficient to yield at least one detection with high probability.
   b) Understand the atmospheric and surface conditions of those exoplanets.
   c) Specifically, search for water and biosignature gases on those exoplanets.
2) Place the Solar System in the context of a diverse set of exoplanets.
   a) Directly detect reflected starlight from a wide range of exoplanets, and transit spectra from a wide range of exoplanets, in order to:
   b) Understand the atmospheric structure and composition of these exoplanets, and
   c) Search for signs of habitability and biological activity in non-Earth-like planets.
   d) Image faint debris disks and exozodiacal light, in order to constrain their structure and composition and lend insights on planet formation processes.
   e) Characterize the architectures of exoplanet systems as a function of stellar type over time.
3) Study and characterize protoplanetary disks.
   a) Image surface brightness features in protoplanetary disks, and determine whether these features as caused by newly-formed planets.
   b) Search for faint scattered light from the material crossing the central, optically-thin clearings in "transitional disks" nearing the ends of their lives.
   c) Determine the locations of the outer edges of protoplanetary disks, in order to constrain how fine dust moves across the region corresponding to the Kuiper Belt in our own planetary system.
   d) Search for vortices within protoplanetary disks in scattered light in the nearest systems, thought to be potential sites of planet formation. Use repeat observations to detect the movements of the features, further constraining the locations and properties of the planets or planet-forming processes responsible.

As outlined in Table 1, the notional design of the HabEx mission includes a UV-capable telescope and instrument suite, with the primary goal of addressing cosmic origins science. However, with this capability, HabEx could also provide important information on the stellar environment of exoplanets, vital for accurate calculations of atmospheric heating and photochemical processes. UV spectroscopic capability in particular is especially important, as these photons play a critical role in photochemistry. Stellar UV flux from late-type main sequence stars is produced by magnetic activity, which is difficult if not impossible to model and cannot be observed with other facilities.

## 2.3 LUVOIR Surveyor

A Large UV/Optical/near-Infrared (LUVOIR) surveyor mission would bring tremendous capabilities for exoplanet science. Possible realizations of such a mission have been previously explored in the NASA Astrophysics Roadmap, and, more recently, in the "From Cosmic Birth to Living Earths" AURA report[11]. We provide a notational range of possible mission architectures for LUVOIR in Table 1. Such a mission would provide incredible science capabilities that are relevant to astrobiology and to comparative (exo)planetology. LUVOIR could have a profound impact on these areas of science by pursuing the following list of objectives:

---

[c] In this document, we describe missions that are capable of searching for habitable conditions and biosignatures. However, we do not specifically define the term "habitable," nor do we specifically define what we consider to be "biosignatures." This is because the definitions of these terms have not yet been standardized across the exoplanet community. In section 6.1.1 we argue that community activities along these lines are ultimately required, in order to properly define these terms and thus the requirements for missions like HabEx and LUVOIR.

4) Search for, characterize, and survey potentially habitable worlds.
    a) Directly detect reflected starlight of Earth-sized planets in the habitable zones of other stars, with a statistically meaningful number of detections, in order to:
    b) Analyze the frequency with which these worlds have certain atmospheric and surface properties, and specifically:
    c) Constrain the frequency of habitability and biological indicators on Earth-sized planets in the habitable zones of other stars.
5) Place the Solar System in the context of a diverse set of exoplanetary systems.
    a) Directly detect reflected starlight from a wide range of exoplanets, and transit spectra from a wide range of exoplanets, in order to:
    b) Understand the atmospheric structure and composition of these exoplanets, and
    c) Search for signs of habitability and biological activity in non-Earth-like planets.
    d) Image faint debris disks and exozodiacal light, in order to constrain their structure and composition and lend insights on planet formation processes.
    e) Characterize the architectures of exoplanet systems as a function of stellar type over time.
6) Study and characterize protoplanetary disks. LUVOIR would also enable the study and characterization of protoplanetary disks, and so address the science goals listed in 3 a-d above.

Because the contrast and resolution requirements for directly imaging potentially habitable planets are generally much more severe than those for imaging of protoplanetary and debris disks, goals 3 and 6 could be achieved be either mission. Of course, because of its (likely) larger collecting area and higher resolution, LUVOIR would enable the study of disks in finer detail at fixed distance, and the study of disks over a larger volume. The UV spectroscopic capability of LUVOIR would allow the characterization of high-energy radiation emission from exoplanet host stars, as described above for HabEx.

Although there is a high degree of overlap with the goals of LUVOIR, the difference in science goals 1) and 4) is significant and worth emphasizing. Our community feels that a mission that can find, image, and characterize potentially habitable exoplanets is worthy of a flagship mission. However, we are explicitly differentiating the expectation to image potentially habitable worlds and *search* for signs of life from the expectation to image enough potentially habitable worlds to *detect* biosignatures. The former only requires that we search for such signals, with no guarantees of detecting anything specific. The latter requires that we search for such signals on a sufficient number of targets to have a reasonable degree of confidence of finding something, or to at least constrain the abundance of such signals to a reasonable level. Goals 4c) and Goal 1c) are consistent with this differentiation. Goal 1c) only requires that we image potentially habitable worlds, obtain their spectra, and analyze their spectra for signs of habitability and life. It makes no assurances with respect to the detection of specific gases. Goal 4c) is significantly more ambitious, and would allow us to begin to estimate the abundance of habitable planets and possible biosignatures, and therefore turn the question of "Are we alone?" into a slightly more quantitative question: "How common are possible biosignatures on Earth-sized planets in the habitable zones of other stars?" And, should the abundance of those signatures be over a certain level (yet to be quantified) in the local galaxy, it would give us a very good expectation of detecting them at least once.

Addressing these science questions with both LUVOIR and HabEx are associated with a diversity of challenges. Some of these are science challenges that will require an interdisciplinary perspective to address. Thus, as we discuss in more detail below, we believe the STDTs should be composed of scientists with a diversity of expertise and significant interdisciplinary perspective to ensure the success of their activities and maximize the viability of this mission. We therefore believe that, the STDTs of both HabEx and LUVOIR should have substantial representation from exoplanet scientists, planetary (solar system) scientists, and astrobiologists.

Other challenges are technical in nature, and we anticipate a need for further investment in technology development to maximize the opportunities for exoplanet-related science with LUVOIR and HabEx-like missions. The STDTs should have an opportunity to identify those areas in greatest need of development in advance of the decadal survey. Further, because many of the technical challenges faced by LUVOIR in particular may have the greatest impact on exoplanet science, most of the risk associated with not overcoming those challenges may be in achieving the maximal exoplanet science return. As a result, there is a desire in the exoplanet community for flagship-scale missions with technical challenges that differ in both scope and in detail, which is the primary reason we advocate studying a range of architectures, including both LUVOIR and HabEx.

## 2.4 The X-Ray Surveyor

While we found that the potential exoplanet science applications of an X-Ray Surveyor were more limited than those of the Far-IR Surveyor and (in particular) HabEx and the LUVOIR Surveyor missions, we nevertheless identified compelling exoplanet science applications of this mission, as discussed below. Possible realizations of such a mission have been previously explored in the NASA Astrophysics Roadmap[3]. We provide a notational range of possible mission architectures for the X-Ray Surveyor in Table 1.

Although the exoplanet environment is traditionally thought of in terms of its relation to the host star's bulk photospheric flux, exoplanets also experience high energy ultraviolet and X-ray radiation, produced in the magnetically heated upper layers of the host star's atmosphere. In the past decade, there has been a growing recognition of the importance of the stellar high-energy radiation environment in determining planet formation, evolution, and possible subsequent habitability. From an exoplanetary perspective, this radiation controls the planets resulting temperature profile, atmospheric mass, and photochemistry. Therefore, understanding the high energy radiation environment in which the planet resides is extremely important for a proper grasp of planetary characteristics, as well as for the interpretation of any observations of atmospheric composition (especially as regards to detection of potential biomarker gases).

In recent years, attention has moved from not only the importance of the ultraviolet emission of exoplanet host stars, but to their X-ray emission. Soft X-rays play a significant role in exoplanetary atmospheric heating and escape; this emission is important for all host stars during stellar youth, when magnetic activity is high, but it is particularly significant for exoplanets orbiting low mass stars (i.e. M dwarfs). M dwarf stars are now known to host small exoplanets in equal or greater proportion as their more massive solar-like cousins[12], and because they are also vastly more common than more massive stars (comprising 70% of all stars in the Milky Way), by number they are therefore the most likely hosts of potentially habitable exoplanets. However, as the habitable zones of M dwarfs lie at small orbital radii, otherwise potentially habitable planets orbiting these stars are particularly vulnerable to the effects of high-energy radiation. In addition, the stellar flare events that produce X-ray emission remain significant for these stars long past their youth, with X-ray and ultraviolet variability remaining high even for stars that do not flare frequently in optical wavelengths[13]. It is not, however, immediately obvious whether these effects will be detrimental or beneficial: while small, Earth-size planets may have their atmospheres eroded completely[14,15], slightly larger planets (i.e. mini-Neptunes) may actually be eroded into volatile rich super Earths. As planet formation through accretion of local material around M dwarfs is expected to produce very dry worlds[16], atmospheric erosion may provide an important alternate pathway to planetary habitability[17].

The exoplanetary motivation for an X-Ray Surveyor mission is multifold: first and foremost, it would provide a means of characterizing the high-energy radiation environment of newly discovered exoplanetary systems. In the case of transiting worlds, observations of the X-ray transit light curve itself can also directly probe the atmospheric escape of exoplanets by allowing a measurement of the planetary exosphere [18], constraining both the rate of escape and likely composition of the escaping material. In addition, an X-Ray Surveyor provides a means for identifying young exoplanet host stars. By targeting these stars in their youth, either for detection of previously undiscovered exoplanets or follow-up of known planets, one can study the earliest stages of planetary formation and evolution. The youth of these systems is important not only for understanding the effects of high-energy radiation on the planetary atmosphere, as well as studying that environment as a function of age[19], but also for identifying young systems that might be used to constrain orbital evolution scenarios.

## 3. Other Flagship Missions Considered: A Terrestrial Planet Interferometer

Ultimately, there is a need to characterize directly imaged exoplanets in the thermal infrared (roughly 5 - 20 microns for temperate planets). Mid-infrared observations are complementary to optical observations for four reasons.

First, thermal measurements of a planet directly constrain its surface temperature, by enabling an estimate of how much energy a planet radiates. Combined with measurements of the incident stellar flux, the infrared and optical observations together provide an estimate of the planet's Bond albedo and internal heating mechanisms.

Second, infrared spectroscopy can be used to place constraints on the presence and abundance of many additional greenhouse gases (e.g. $CO_2$, water), which absorb thermal infrared radiation and strongly influence the planet's surface temperature and habitability. A planet's Bond albedo and greenhouse gas abundance are the most important determinants of the planet's climate, including its potential habitability.

Third, time-resolved thermal measurements constrain a planet's response to variable incident stellar flux. The planet's rotation causes diurnal variation, while its obliquity and eccentricity may cause seasonal variation. The response to varying radiation forcing constrains the thermal response time of the planet, and hence the atmospheric mass, the presence of a heat sink (e.g. an ocean), and heat transport efficiency.

Finally, mid-infrared spectroscopy is sensitive to many known biosignature gases, including $O_3$ and $CH_4$ (but not $O_2$, which is detectable in the optical). Observation of some of these gases will likely be needed to confirm that any biosignatures detected in the optical with LUVOIR or HabEx are true signs of extrasolar life.

Unfortunately, measuring the thermal emission of an Earth analog exoplanet is extremely challenging. Obtaining the required angular resolution at these wavelengths demands interferometry; obtaining the needed extreme contrast requires suppressing starlight while outside Earth's atmosphere (e.g. nulling interferometry, as used by the Darwin[20] and TPF-I[21] mission concepts). Neither of these technologies is yet mature for space-based platforms, and thus they warrant further study and development. Further discussion of this topic appears in the section on the Far-IR Surveyor (Section 2.1).

## 4. Characterization of Transiting Planets: Prospects for LUVOIR, HabEx, and a Probe-Class Mission

The primary method of characterizing exoplanets used by the LUVOIR and HabEx missions described above is direct imaging and spectroscopy. Yet, the current workhorse method of characterizing exoplanets is entirely different: transit spectroscopy. The transit spectroscopy method exploits the spectral-temporal modulation of light from a star-planet system that occurs in cases where the plane of the planet's orbit causes the planet to pass in front of, or behind, the parent star. Measurements of these events can be used to infer the emission and/or reflection properties of a planet's atmosphere and surface as well as the transmission spectrum of the planet's atmosphere. Transit spectroscopy, both in the visible and near infrared, has been one of the only methods for studying the composition and conditions of exoplanet atmospheres. Current transit characterization capabilities are primarily limited to the Hubble Space Telescope and Spitzer Space Telescope, and ground-based facilities in some exceptional cases. Only about 20 exoplanets have been characterized in detail over the last five years with these resources, because most of the known transiting planets orbit faint stars. Therefore, our understanding of the atmospheres, compositions, and properties of transiting planets (and therefore, exoplanets in general) remains poor.

The combination of the Transiting Exoplanet Survey Satellite (TESS)[22] and the PLAnetary Transits and Oscillations of stars (PLATO)[23] missions will find thousands of bright transiting planets, ranging from gas giants to temperate terrestrials, providing a wealth of potential targets. The TESS discoveries, in particular, will enable an atmospheric survey of $10^2$ to $10^3$ bright hot Jupiters and warm sub-Neptunes using JWST. Planets are extremely complex: the mass and temperature of a planet provides exceedingly poor constraints on a planet's atmospheric composition, dominant chemical processes, evolutionary history, and formation scenario. The overarching questions of exoplanetary diversity, formation, and evolution all require, or are assisted by, the detailed study of a large and diverse set of exoplanetary atmospheres and surfaces.

TESS is also expected to discover a few temperate terrestrial planets transiting nearby M-Dwarfs, which may host habitable conditions or even life. Additional complexity is introduced by our quest to identify potentially habitable worlds outside our solar system where life may exist. Properly characterizing these "temperate terrestrials" identified by TESS will be time-intensive: JWST will need months to provide tantalizing constraints on the presence of an atmosphere, planetary rotational state, clouds, and greenhouse gases.

As argued in the ExoPAG SAG10 report "Characterizing Transiting Planet Atmospheres through 2025"[24], even assuming optimistic allocations of time on JWST for transit characterization, it is unlikely that JWST will be able to carry out both a comprehensive atmospheric survey of a large number of hot Jupiters and warm sub-Neptunes *and* the detailed characterization of the temperate terrestrial planets discovered by TESS. Furthermore, depending on several currently unknown factors, it is unclear if JWST will be able to definitively demonstrate habitable conditions on these temperate terrestrial planets. Finally, given its predicted launch date of 2024, most of the planets detected by PLATO will be discovered after the nominal mission lifetime of JWST.

Thus, in planning for the 2020 decadal survey, we must consider the post-JWST era and what role transit spectroscopy may have in the study of exoplanets. Clearly, transit spectroscopy has a potentially critical role, and is a needed capability, beyond the operational JWST timeframe.

Given the questions of diversity, habitability, formation, and evolution, we can anticipate some aspects of transit-spectroscopy-capable missions that would have high scientific values in the post JWST era. These would include:
1. Infrared spectroscopic capability to detect and determine the abundance of the important molecules in exoplanet atmospheres.
2. Visible spectroscopic capability to determine the amount of aerosols and clouds in exoplanet atmospheres.

3. UV spectroscopic capability to determine the level of photochemical activity likely present in an exoplanet's atmosphere.
4. Simultaneous visible and infrared spectral coverage to reduce uncertainty introduced by stellar variability.
5. Simultaneous UV-visible-IR spectroscopic capability to reduce uncertainty introduced by stellar variability.
6. Capability to observe many hundreds or thousands of planets.

For the Astronomy 2020 Decadal Survey, there are currently three mission possibilities that can be considered in light of the overarching science questions and measurement capability for transit spectroscopy. A short notional description of each concept is given below.

**HabEX** In addition to the instrument required for the direct detection and characterization of exoplanets, a UV imaging and spectroscopic instrument will likely enable a broad range of cosmic origins science. When combined with a visible-light instrument with a wavelength range out to 2.5 microns and R~200 spectroscopic capability, the UV and optical instruments would allow for valuable transit characterization measurements. Depending on the aperture and precise instrumentation of HabEx, it may be able to follow-up the most promising but difficult targets identified by TESS and JWST, and improving the characterization of their atmospheres.

**LUVOIR** is a general astrophysics flagship mission concept based on a warm 8-16m telescope with multiple instruments covering a broad range of astrophysical capabilities (see Table 1). Whether infrared capability between 2.5 microns and 5.0 microns is included is unclear; coverage to 5 microns is very important for biomarker false positive rejection. UV and visible imaging spectroscopy capability is assumed, but, similar to Hubble, it is not clear that UV+visible+IR capability can all operate on the same target simultaneously. Nevertheless, for apertures on the larger side of those nominally considered for LUVOIR, it will likely be able to better characterize the most promising temperate terrestrial planets identified by JWST, and thus most definitively assess whether the conditions on these worlds are indeed habitable.

**A Probe-class Transit Spectroscopy Mission** represents a concept for a transit spectroscopy survey mission based on a relatively small-aperture, passively cooled ~1.5 m telescope. In order to be optimized for a transit characterization, such a mission should have the capability of continuous and simultaneous spectroscopy in the UV, visible, and IR for wavelengths as long as 8 microns. Such a mission operating for 5 years, with the majority of time dedicated to transit spectroscopy, could characterize a large number (several hundreds) of transiting hot Jupiters and warm Neptunes, although its aperture would be too small to enable the characterization of temperate terrestrial planets.

## 5. Probe-class Missions

There was a general consensus in the ExoPAG that there exist compelling probe-class (<$1B) missions that could contribute significantly to exoplanet science. We identified three missions that we advocate should be considered for detailed study in advance of the next decadal survey, both because of the science capabilities they enable, and because they may represent possible exemplars of compelling probe-class missions that could be used to justify a dedicated probe-class mission line, analogous to the New Frontiers program in NASA's Planetary Science Division.

## 5.1 Starshade for WFIRST-AFTA

A probe class starshade mission can rendezvous with and effectively leverage WFIRST-AFTA to capture early spectra from Earth-like exoplanets and critically inform the design of future exoplanet flagship missions. Continuing dark energy observations in parallel with starshade observations minimizes the impact to primary mission objectives. WFIRST-AFTA can be made starshade ready with minor modifications to the baseline coronagraph instrument and by adding a radio system for starshade communications and range measurement. The case study Rendezvous Mission detailed in the Exo-S STDT Report[10] is a 3-year Class C[25] mission that targets a broad range of planet types and emphasizes low-cost and technology readiness over science performance. Another attractive scenario to consider is a 5-year Class B mission optimized to detect Earth-like exoplanets. Striking an optimal balance between the Inner Working Angle (IWA) and bandwidth opens up access to a large number of habitable zones. Adding solar electric propulsion and mission time enables their exploration. Early simulations show that at least 10 exo-Earths can be detected if $\eta_{Earth}$ is at least 20%. A much larger number of other planet types will also be detected. The option remains to spectrally characterize a subset of detected planets, over a wider bandwidth and larger IWA.

We note that the motivation for a starshade "rendezvous" mission rests upon the assumption that the WFIRST will remain the next NASA flagship mission after JWST, and that it will include an internal coronagraph instrument. Should either (or both) assumptions end up being violated, than probe-class direct imaging missions similar to the Exo-C and/or Exo-S mission concepts should be considered in more detail, as they could provide needed demonstration of starlight suppression technologies in flight.

## 5.2 Transit Characterization Mission

As discussed in Section 3 above, a dedicated transit spectroscopy survey mission based on a passively cooled ~1.5 m telescope would provide a highly-capable mission for characterizing the atmospheres of a large number of transiting exoplanets. The basic mission concept would include continuous and simultaneous UV, visible, and IR spectroscopy for wavelengths up to 8 microns, with a mission lifetime of 5 years. Such a mission would enable a large survey of exoplanets. Whether or not a probe-class transit characterization mission is justified will depend somewhat on amount of time for transit characterization available on JWST. A detailed discussion about the scientific justification for a probe-class mission in concert with JWST is provided in the SAG 10 Report[24].

Nevertheless, at least for large planets orbiting bright stars, such a probe-class mission would likely prove substantially more capable than JWST, HabEx or LUVOIR, due to both the ability to acquire simultaneous spectroscopy to longer wavelengths, and due to the much larger amount of time that would be spent on transit characterization observations. Note that such a probe-class transit characterization mission would *not* be capable of detailed characterization of temperate terrestrial planets. These measurements require JWST at the very least, and perhaps additional instrumentation on HabEx or LUVOIR, as described above. Furthermore, since the probe-class mission would almost certainly be launched well after JWST, it is unlikely to be able to identify interesting systems for further study with JWST.

Thus, depending on the returns of JWST, it may be that the best way to maximize the scientific returns of TESS and PLATO is a "three-pronged" approach. JWST could be used to perform detailed atmospheric characterization of the most interesting transiting targets identified from TESS (transit, eclipse, and - when possible - phase-resolved spectroscopy), including the handful of temperate terrestrial planets to search for potentially habitable conditions. Instruments on a flagship mission like HabEx or LUVOIR may then allow for better constraints on the habitability of these terrestrial planets. Provided the community still identifies the need, a probe-class mission operating in the next decade with the notional parameters described above could perform a comprehensive survey of a large fraction of the bright Hot Jupiters and Warm Neptunes identified by TESS and PLATO, measuring eclipses and phase variations for several hundred planets.

### 5.3 Astrometry Mission

As an example of a probe-class astrometry mission, we consider a ~1.2m astrometric telescope, with a 0.25 deg$^2$ FOV, which would use novel technologies to control systematic errors to near photon-limited performance. An example of such a mission is the ESA Theia[d] concept[26], a 0.8m astrometric telescope with an estimated cost of ~ $630M. A more ambitious 1.2 m telescopes would still likely fall below the $1B cost cap of a probe class mission. As the photon-limited astrometry integration time goes as $D^4$, a 1.2m telescope could observe nearly twice as many targets at 1 µas precision than a 1m telescope. For bright targets (i.e., nearby stars being searched for Earth analogs) the accuracy is limited by photon noise of the reference stars. For faint targets the photon noise of the target is the limiting factor. The photon limited accuracy for stars V< 7 mag ~0.4 µas in 1 hr. At V~10 the accuracy is ~1.0 µas in 1 hr, and is limited by the photon noise of the target. For V~15, the photon-noise precision would be ~10 µas in 1hr. Note that the achievable precision of such a mission would be considerably better than that expected for Gaia[27], which is expected to achieve an *end-of-mission* measurement precision of at best ~10 µas for stars brighter than V~12.

Such a probe-class astrometry mission would enable the detection of Earth-mass planets around the stars nearest to the Sun. A Sun-Earth clone at 10 pc would have an astrometric signature of 0.3 mas. Assuming a SNR~6 for detection (< 1% false alarm probability), for bright (V<7) stars, a total integration time of 64 hours (spread over > 20 epochs over 5 years) would enable the detection of an Earth analog at a signal-to-noise ratio of ~6 (< 1% false alarm probability). Assuming ~25% of a 5 year mission is devoted to searching the most nearby stars for Earth analogs, ~160 stars can be surveyed down one Earth mass in the middle of the habitable zone out to a distance of 20 pc. If $\eta_{Earth}$ is 10%, 16 Earth analogs would be discovered. Other planets with larger astrometric signatures would also be detected, including most of the known RV planets, for which the orbital inclination would be measured.

In principle, such a survey could identify targets for a HabEx or LUVOIR mission, thereby potentially increasing the efficiency of these missions and/or enabling the same science yield for smaller apertures. The utility and value of an astrometry mission to act as such a precursor mission depends on many factors, such as the achievable astrometric precision, the architecture and starlight suppression technology of HabEx or LUVOIR, the sample of stars being targeted, and the frequency of potentially habitable planets, and others. Therefore, this potential application should be studied in more detail. Regardless, such a mission would likely be able to measure the masses and refine the orbits of planets imaged by HabEx or LUVOIR, which is crucial for interpreting the habitability of these planets, and may be difficult or impossible with any other method.

## 6. Concerns Regarding the Structure and Charge of the STDTs

### 6.1 Concerns Specific to the HabEx and LUVOIR Missions

#### 6.1.1 The Need for Astrobiological and Biosignature Standards

This document lists a search for biosignatures as a goal and science driver for both LUVOIR and HabEx. However, the definitions of terms such as "biosignature,"

---

[d] Not to be confused with the Telescope for Habitable Exoplanets and Interstellar/Intergalactic Astronomy (THEIA) mission[8] discussed in Section 2.2.

"habitability," "Earth-like," and "signs of life" have not been standardized across the exoplanet community. This is a fundamental limitation on our ability to do such science, as these definitions should have some standardization prior to implementation of requirements for positive detections for these missions. There is therefore a need to incorporate lessons from the astrobiology community – which has been searching for signs of life on other planets in our solar system, and in the deep time history of our home planet – into the STDTs for these missions. In addition to informing definitions of these astrobiological terms, this will also allow better definition of what constitutes a biosignature, on discriminating between living planets and those with false positives for life, and on expanding the suite of biosignature gases the STDTs would be aware of. Likewise, it is critical to incorporate the expertise of exoplanet scientists that fully understand how past observations have upset expectations based on observations of our own solar system, and of models of exoplanetary systems. Ultimately, this will allow for an analysis of the extent to which either of these missions is capable of conducting a search for life, and an assessment of the technical capabilities required to conduct such a search. It will also allow these STDTs to be as thorough as possible in predicting the observations of categories of planets for which we have few – or in some cases zero – examples on which to base our expectations. This includes a comprehensive examination of "false positive" or "false negative" planets that respectively exhibit signs of life despite the lack of a biosphere, or exhibit no signs of life despite being inhabited.

### 6.1.2 Topical Representation on, and Coordination between, LUVOIR and HabEx STDTs

Regarding the Habitable-Exoplanet Imaging Mission (HabEx), and the Large UV/Optical/IR Surveyor (LUVOIR), the ExoPAG recommends that the direct detection and spectroscopic characterization of habitable exo-Earths be one of the highest priority science goals for these missions, and that the overall mission designs be compatible with that key science objective. For LUVOIR in particular, we advocate that this requirement be obtainable for statistically significant number of habitable exo-Earths. We also emphasize that knowledge drawn from Solar System observations is crucial to the definition of mission science requirements and data interpretation. Likewise, NASA mission science goals directly and indirectly benefit from supporting observations from ground-based facilities.

Similarly we feel that, in order to ensure broad support for both missions within both the exoplanet and cosmic origins communities, that significant cosmic origins (and exoplanet) science capabilities be baselined for *both* the HabEx and LUVOIR missions.

We therefore believe it is vital to ensure sufficient representation of the relevant scientific and technical expertise on both the HabEx and LUVOIR STDTs. We recommend that the membership of these STDTs reflect this high and equitable priority for both cosmic origins and exoplanet science. This can be accomplished by drawing a roughly equal fraction of the members of each STDT from the broad exoplanet community (including planetary scientists and exobiologists) on the one hand, and the cosmic origins community on the other hand.

We further suggest that the two STDTs should hold joint meetings, especially during the early stages. In particular, we recommend that they coordinate their starting assumptions on the state of the field, and that significant cross-talk take place during the detailed drafting of their science cases. Having common members of both teams (if feasible) may facilitate such coordination.

### 6.1.3 Standards Definition and Evaluation Teams

We argue that NASA should consider setting up a small, transparent and unbiased

"standards" team, in addition to the primary STDTs for each mission. The primary charge of this team would to evaluate the science yield and technical readiness of all the mission designs in a consistent and transparent manner, using agreed-upon sets of assumptions and methodologies. We note that creating one such team is likely to be particularly important for comparisons between HabEx and LUVOIR, as both the science goals and technological requirements of these missions are likely to have significant overlap, and NASA, as well as the community as a whole, will ultimately need to be compare and evaluate the expected science yields and technological requirements of these two missions in a consistent and fair manner. We note that this is one point of partial disagreement between the ExoPAG and COPAG reports. In particular, the COPAG states that: "There is no compelling reason to set up an independent review team outside of the STDTs to assess the scientific integrity of the STDTs' Cosmic Origins science assumptions or technical requirements, as is being recommended by the ExoPAG for the characterization of Earth-like exoplanets."

## 6.2 General Concerns

### 6.2.1 Costing

There is a clear need for a *fair*, *realistic*, and *consistent* cost analysis of flagship missions, done in advance of the decadal survey process, that is carried out on architectures that are *responsive* to the top-level science goals above. This will require a costing process that is applied equally to all STDTs (fairness), that best leverages the expertise of qualified, independent experts (realism), that attempts to level different estimates through interactions of different costing teams (consistent), and that measures cost relative to specific science requirements that are achieved by an architecture (responsiveness). We specifically note that the opportunity for interactions between the STDTs and the Cost Analysis and Technical Evaluation groups for the two probe studies (Exo-S and Exo-C) was considered useful, and that an opportunity for such interactions provides the community with a good model for the assessment of future missions.

### 6.2.1 International Representation

A large mission for exoplanet science (LUVOIR or HabEx) is very likely to have a significant international component, reflecting both the technical scale of the mission but also the breadth of its science appeal. International partnerships in a large mission will require discussion and planning over an extended period. The STDTs that will be developing these mission concepts provide an opportunity to begin that dialog.

We recommend that NASA appoint one or two international 'Observers' to each STDT. Observers would not be full members. Instead, they would be invited to attend, but not directly participate in, STDT meetings and telecons, and to receive copies of materials discussed during the meetings. This would ensure that potential future partners are accurately informed of developments within NASA. We recognize two important restrictions on their participation. First, a key STDT objective is to provide input to the 2020 US Decadal Survey, and formal international participation should not be part of the process. Second, restrictions may need to be imposed to meet ITAR/EAR requirements.

We feel that this simple step would help lay the groundwork for future collaborations should one of the large missions be recommended by the 2020 Decadal Survey.

### 6.2.3 Representation From Scientists in "Soft Money" Positions

The United States is unique in its strong support for independent science organizations and a community of highly experienced, independent, "soft money" scientists. In forming the STDTs, it is strongly advisable to seek and enable participation from the most qualified exoplanet scientists, including soft money and/or non-tenured researchers who may not have institutional resources for extensive pro bono contributions. This could be achieved by providing a reasonable level support for STDT members to devote 10-20% of their time to the effort, depending on current and pending funding requirements and expected contributions. Strictly pro bono participation discourages broad segments of the astronomical community, especially younger researchers and independent scientists, who could otherwise offer fresh insight, useful tools, and valuable decision-making.

## 7. Conclusions and Summary

Through a series of face-to-face and virtual meetings, the ExoPAG discussed at length its planned response to NASA's charge to identify a small (3-4) set of large mission concepts to be studied by Science and Technology Definition Teams (STDTs) in advance of, and in preparation for, the 2020 Astrophysics decadal survey. The consensus of the ExoPAG developed from these discussions is a recommendation that all four large mission concepts identified in the white paper as candidates for mission concept maturation prior to the 2020 Decadal Survey should be studied in detail. These include the Far-IR Surveyor, the Habitable-Exoplanet Imaging Mission, the Large UV/Optical/IR Surveyor, and the X-ray Surveyor. ExoPAG found that all four proposed mission concepts could address, to varying extents, important exoplanet science goals of the next 10-20 years. These science applications are outlined in Section 2 of this report.

The ExoPAG considered other flagship mission concepts for study, but none achieved sufficiently broad community support to be elevated to the level of these four primary candidate missions. Nevertheless, we concluded that at least one of the missions considered, namely a mid-infrared interferometer that can directly image the thermal emission of Earthlike planets, is likely to eventually be required to confirm the habitability of these worlds. Such a mission is extremely challenging, and the two most demanding required technologies (space-based high-contrast interferometry and formation flying) remain immature. We therefore advocate sustained investment in these technologies.

The ExoPAG identified a potential need for additional capabilities to characterize transiting exoplanets beyond those that will be available with JWST. This need may be addressed either by instruments on HabEx or LUVOIR, or via a dedicated probe-class mission, or both.

There was a general consensus that there exist compelling probe-class (<$1B) missions that could contribute significantly to exoplanet science. We identified three missions that we advocate should be considered for detailed study in advance of the next decadal survey: a starshade for WFIRST-AFTA, a dedicated transit characterization mission, and an astrometry mission. These mission are potentially compelling, both because of the science capabilities they enable, and because they may represent possible exemplars of viable and compelling probe-class missions that could be used to justify a dedicated probe-class mission line.

Finally, the ExoPAG outlined several concerns about the structure and charge of the STDTs that will be assembled in response to these reports. In particular, we identified a need for a fair, realistic, and consistent cost analysis of the missions, suggested that the representation of the LUVOIR and HabEx STDTs not be too heavily weighted to any particular discipline, and include representation from the planetary science community, as well as participation from the international community. We note that scientists in "soft money" positions, whose expertise is likely to be important for the success of the STDT activities, are at a disadvantage relative to those with more permanent funding lines. Therefore, financial support of these participates should be considered to ensure the participation of the most qualified scientists. Finally, we suggested that NASA consider assembling a transparent and unbiased "standards" team, in addition to the primary STDTs for each mission; the charge of this team would be to evaluate the science yield and technical readiness of all the mission designs in a consistent and transparent manner.

## Acknowledgements

The ExoPAG would like to thank Paul Hertz for giving us opportunity to provide input on the important decision of which large missions should be studied in detail in advance of the decadal survey. We would like to thank all of the members of the astronomical community, and the exoplanet community in particular, who provided input to this report. We acknowledge the extensive activities of the COPAG and PhysPAG in responding to this charge, and single out the efforts of Ken Sembach (COPAG EC Chair) and Jamie Bock (PhysPAG EC Chair) in particular, for their help with coordinating the process of gathering community input, and forging a consensus amongst the three PAGs.

## Appendix A: Processes and procedures used to solicit and incorporate community response

The ExoPAG had already initiated the process of building consensus for an "Exoplanet Roadmap" through the SIG1 activities. The SIG1 was approved during the March 26-27, 2014 meeting of the NASA Advisory Council Astrophysics Subcommittee meeting. The charge of the SIG1 is as follows:

> ExoPAG SIG #1: Toward a Near-Term Exoplanet Community Plan
>
> The goal of this Science Interest Group is to begin the process of developing a holistic, broad, unified, and coherent plan for exoplanet exploration, focusing on areas where NASA can contribute. To accomplish this goal, the SIG will work with the ExoPAG to collect community input on the objectives and priorities for the study of exoplanets. Using this input, it will attempt to develop a near-term (5-10 year) plan for exoplanets, based on the broadest possible community consensus. The results of this effort will serve as input to more formal strategic planning activities that we expect will be initiated near the end of the decade in advance of the next decadal survey.

| | |
|---|---|
| 1/5/2014 | ExoPAG 9, Washington, DC |
| 6/6/2014 | ExoPAG 10, Boston, MA |
| 1/4/2015 | ExoPAG 11, Seattle, WA |
| 2/10/2015-2/11/2015 | SIG1 Stand-alone Meeting, Pasadena, CA |
| 6/2/2015 | SIG1 Virtual Meeting #1 |
| 6/14/2015 | ExoPAG 12, Chicago, IL |
| 7/14/2015 | SIG1 Virtual Meeting #2 |
| 8/18/2015 | SIG 1 Virtual Meeting #3 |

Table 2: Meetings in which the SIG1 and the "Charge to the Astrophysics PAGs regarding Large Mission Concepts" were discussed by the ExoPAG. These meetings were typically attended by ~40-50 people (in person and/or remotely). These attendees represented a reasonably broad cross section of the exoplanet community, as well as some representation from the Cosmic Origins and Physics of the Cosmos communities.

The ExoPAG has been working to respond to NASA's charge under the auspices of this SIG. The primary process by which we solicited community input was through face-to-face and virtual meetings. These meetings were also used to generate consensus points, the most important of which are summarized in Appendix B. These points formed the basis of the report. The ExoPAG-specific meetings in which at least part of the time and discussion were devoted to responding to the charge are listed in Table 3. In addition, several members of the ExoPAG attended relevant meetings of the COPAG, PhysPAG, and their SIGs. Finally, we drew input from the COPAG-solicited white papers, which can be found on the [COPAG website](#) devoted to responding to the charge, as well as various reports, including the NASA Astrophysics Roadmap[3], the TPF-C Report[28], the New Worlds Observer report[29], the THEIA report[8], the Exo-C[9], and Exo-S[10], reports, and finally the AURA HDST report[11].

## Appendix B: Points of Consensus Achieved During ExoPAG 12

The second day of the ExoPAG 12 meeting in Chicago, IL, was devoted to talks related to, and discussions about, the charge addressed in this report. One outcome of these discussions was a set of seven points of consensus among the attendees of this meeting. Specifically, the following propositions were put in front of the attendees, and a vote was taken (via a show of hands) as to whether or not the attendees agreed with the propositions. In all cases, all those in attendance who chose to vote, which constituted the majority of the attendees, supported the propositions below.

1. There was a general support for the WFIRST mission with a coronagraph **and** a starshade.
2. There was a general consensus that a broad range of apertures and architectures for direct imaging missions should be studied, encompassing both the nominal concepts of the HabEx and LUVOIR missions.
3. There was a general consensus that there should be a common executive summary with the other PAG reports. It was agreed that the executive summary should include: a statement that we support these four missions being studied, a recommendation for probe studies, and suggestions for how STDTs should be organized (provided that the other PAGs are in agreement on these points).
4. There was a general consensus that a common table describing the nominal parameters of the four missions should be included in the PAG reports. These parameters are to be determined in future discussions with the ExoPAG and other PAGs[e].
5. There was a general consensus that we should neither add nor subtract from the four proposed mission concepts (HabEx, LUVOIR, X-ray Surveyor, and Far-IR Surveyor).
6. With regards to organization of the HabEx and LUVOIR STDTs, there was a general consensus on the following points:
   a. There should be two separate science teams and two separate engineering and technology teams.
   b. The science teams should have significant overlap (common members), and should include significant representation from the planetary science community.
   c. We should express the following concerns in the report:
      a. Exoplanets may get marginalized in the LUVOIR STDT if their representation is too small.

---

[e] We note that the COPAG subsequently decided not to include such a table in their final report, although the PhysPAG has included this table.

Large Mission Concepts for Study                19/22                6 October, 2015

b. The general astronomical community may get fractured if the representation of disciplines is very different between the two STDTs.
    d. Thus the members of the science teams should be carefully chosen to ameliorate these concerns.
    e. The teams should meet periodically, including the kickoff meeting.
    f. There should be a small, independent and unbiased team that is tasked to evaluate the science yield and technical readiness of both mission designs in a consistent and transparent manner[f].
7. There was a general consensus that probe-class (<~$1B) missions should be studied in advance of the next decadal survey, and that the following missions should be presented in the report as examples of possibly compelling probe-class missions.
    a. A starshade for WFIRST-AFTA.
    b. A transit characterization mission.
    c. An astrometry mission.

# References

[1]Hertz, P. 2015, "Planning for the 2020 Decadal Survey: An Astrophysics Division White Paper."
http://science.nasa.gov/media/medialibrary/2015/01/02/White_Paper_-_Planning_for_the_2020_Decadal_Survey.pdf

[2]Blanford, R., et al. 2010, "New Worlds, New Horizons in Astronomy and Astrophysics."
http://www.nap.edu/catalog/12951/new-worlds-new-horizons-in-astronomy-and-astrophysics

[3]Kouveliotou, C., et al. 2013, "Enduring Quests, Daring Visions: NASA Astrophysics in the Next Three Decades."
http://science.nasa.gov/media/medialibrary/2013/12/20/secure-Astrophysics_Roadmap_2013.pdf

[4] http://www.jwst.nasa.gov

[5] Spergel, D.N., et al., 2015, "Wide-Field InfraRed Survey Telescope-Astrophysics Focused Telescope Assets (WFIRST-AFTA) 2015 Report."
http://wfirst.gsfc.nasa.gov/science/sdt_public/WFIRST-AFTA_SDT_Report_150310_Final.pdf, http://wfirst.gsfc.nasa.gov

[6]NASA 2014 Science Plan.
http://science.nasa.gov/media/medialibrary/2014/05/02/2014_Science_Plan-0501_tagged.pdf

---

[f] We note that the COPAG disagreed with this conclusion, i.e., they did not feel that such a "standards" team was needed.

[7] Armus, L., et al 2015, "From Early Galaxies to Habitable Planets: The Science Case and Concept for a Far-Infrared Surveyor."
http://conference.ipac.caltech.edu/firsurveyor/system/media_files/binaries/49/original/FIRsurveyor_galaxies-to-planets.pdf?1439406838

[8] Kasdin, N.J., et al. 2009, "THEIA: Telescope for Habitable Exoplanets and Interstellar/Intergalactic Astronomy."
http://www.princeton.edu/~hcil/papers/theiaWhitePaper.pdf

[9] Stapelfeldt, K.R., et al. 2015, "Exo-C: a probe-scale space mission to directly image and spectroscopically characterize exoplanetary systems using an internal coronagraph."
https://exep.jpl.nasa.gov/stdt/Exo-C_Final_Report_for_Unlimited_Release_150323.pdf

[10] Seager, Sara, et al, 2015, "Exo-S: Starshade Probe-Class Exoplanet Direct Imaging Mission Concept, Final Report."
https://exep.jpl.nasa.gov/stdt/Exo-S_Starshade_Probe_Class_Final_Report_150312_URS250118.pdf

[11] Dalcanton, J., et al. 2015, "From Cosmic Birth to Living Earths: The Future of UVOIR Space Astronomy."
http://www.hdstvision.org/s/hdst_report_final_072715.pdf

[12] Dressing, C. D. et al. 2015, ApJ, 807,45, "The Occurrence of Potentially Habitable Planets Orbiting M Dwarfs Estimated from the Full Kepler Dataset and an Empirical Measurement of the Detection Sensitivity."
http://adsabs.harvard.edu/abs/2015ApJ...807...45D

[13] France, K., et al, 2015, IAU General Assembly, Meeting #29, "Ultraviolet and X-ray Activity and Flaring on Low-Mass Exoplanet Host Stars."
http://adsabs.harvard.edu/abs/2015IAUGA..2228599F

[14] Lammer, H. et al., 2008, ASPC, 450, "The Loss of Nitrogen-rich Atmospheres from Earth-like Exoplanets within M-star Habitable Zones."
http://adsabs.harvard.edu/abs/2011ASPC..450..139L

[15] Lammer, H., et al, 2013, MNRAS, 430, 1247, "Probing the blow-off criteria of hydrogen rich super Earths."
http://adsabs.harvard.edu/abs/2013MNRAS.430.1247L

[16] Raymond, S. et al., 2007, ApJ 669, 606, "A Decreased Probability of Habitable Planet Formation around Low-Mass Stars."
http://adsabs.harvard.edu/abs/2007ApJ...669..606R

[17] Luger, R. et al, 2015, Astrobiology, 15, 57, "Habitable Evaporated Cores: Transforming Mini-Neptunes into Super-Earths in the Habitable Zones of M Dwarfs."
http://adsabs.harvard.edu/abs/2015AsBio..15...57L

[18] Poppenhaeger, K., et al., 2015, IAU General Assembly, Meeting #29, "A tale of two exoplanets: X-ray studies of the Hot Jupiters HD 189733 b and CoRoT-2."
http://adsabs.harvard.edu/abs/2015IAUGA..2257334P

[19] Tian, F. 2015, Icarus, 258, 50, "Observations of exoplanets in time-evolving habitable zones of pre-main-sequence M dwarfs."

Large Mission Concepts for Study                21/22                6 October, 2015